\documentclass[aps,pra,showpacs]{revtex4} 

\usepackage{amssymb} 
\usepackage{amsmath} 
\usepackage{graphicx}

\begin{document}

\title{Hirota method for oblique solitons in two-dimensional supersonic nonlinear Schr\"odinger flow}

\author{E.G. Khamis}
\email{egkhamis@if.usp.br}
\author{A. Gammal}
\email{gammal@if.usp.br}

\affiliation{ Instituto de F\'{\i}sica, Universidade de S\~{a}o Paulo, 05508-090, S\~{a}o Paulo, 
Brazil}

\date{\today}

\begin{abstract} In a previous work\cite{egk-06} exact stable oblique soliton solutions were 
revealed in two dimensional nonlinear Schr\"odinger flow. In this work we show that single 
soliton solution can be expressed within the Hirota bilinear formalism. An attempt to build 
two-soliton solutions shows that the system is ``close" to integrability provided that the angle 
between the solitons is small and/or we are in the hypersonic limit.

\end{abstract}


\maketitle

1. {\it Introduction} \newline The nonlinear Schr\"odinger (NLS) flow is ubiquitous in many 
physical systems such as photorefractive crystals, and the superfluids Bose-Einstein condensates 
and exciton-polaritons. A fundamental problem is how a superfluid reacts to the presence of an 
obstacle. One can define a Mach velocity $(M)$ as the velocity of the obstacle relative to the 
sound velocity in the medium. Two-dimensional studies showed that when $M>0.37$ the system loses 
superfluidity and start to emit pair of vortices \cite{frisch,berloff2000,rica2001,smirnov}. Increasing the velocity 
showed that vortices merge in a ``vortex street" \cite{winiecki}, which were later understood 
as oblique solitons \cite{ek-06} and its exact single soliton solution determined \cite{egk-06,kgegk-08}. 
Oblique 
solitons were long know to be unstable but it was showed that under the flow they are only  
convectively unstable provided that $M>1.44$ \cite{kp-08,kk-11}. Studies with extended obstacles also 
presented oblique solitons in the wake, and an analytical approach based on Whitham modulation 
theory was successfully applied \cite{ekkag-09}. Oblique solitons were realized 
experimentally in the system of exciton-polaritons \cite{amo-2010}, 
though for lower Mach number than originally predicted by theory. Corrections to the model 
including losses were able to match experimental observations \cite{korneev2011}. 
Dynamics of formation and decay of oblique solitons were recently observed in \cite{grosso2011}.

A key question about solitons is how they behave in collisions. As long as we know, 
there is no general proof about the non-integrability of the 2D-NLS. 
 Numerical studies 
with two obstacles were able to generate collisions between these oblique solitons 
\cite{annibale2011}. These collisions were shown to be practically elastic suggesting 
integrability or ``close" to integrability in such system. In the same work, an analytical 
treatment was considered using hydrodynamical approach and the system was show to follow a 
1D-NLS equation in the hypersonic limit, and collisions could be described by the well known 
phase shifts \cite{zs-1973}. Numerical calculations were in good agreement with predicted phase 
shifts, considering that they were perturbed by previous interactions with linear waves. Since 
the exact single soliton (1SS) was already obtained, one might ask if an exact two-soliton 
solution (2SS) could be found. A possible framework to find multiple soliton solutions is the 
Hirota method \cite{hirota73,hirotabook}. In the following we build up a bilinear Hirota form of 
the 2D-NLS in the stationary frame relative to the obstacle. Then, we show that the single 
oblique soliton solution indeed satisfy this form. In the sequence we propose an ansatz to the 
exact solution of the two-soliton interaction problem and analyze its consequences within this 
formalism.

\bigskip

2. {\it Model} \newline Oblique dark solitons in a superfluid are described 
\cite{egk-06} as stationary solutions of the defocusing nonlinear Schr\"odinger equation (NLS)  
\begin{equation}\label{1-1}
    i\psi_t=-\frac{1}{2} \Delta\psi+|\psi|^2\psi +V(x+Mt,y)\psi\,, \end{equation} which is 
written here in standard dimensionless units, $\Delta \equiv \partial^2_x+\partial^2_y $, 
the subscripts mean derivatives 
and the potential $V$ is modeled as a small impenetrable obstacle. The potential $V$ is moved 
with Mach velocity $M$ from right
to left  across the fluid, where $M$ is in units of the sound velocity. 
We make a global phase transformation $\psi'=e^{it}\psi$ and later a Galilean transformation 
$x'=x+Mt$, $t'=t$ leading to
\begin{equation} 
-2i\psi_t=\psi_{xx}+\psi_{yy}+2iM\psi_{x}+2\psi-2\psi|\psi|^2 -2V(x,y)\psi\,, \label{galileo} 
\end{equation} where the primes were omitted for convenience.

This last equation describes the wave function in the stationary frame relative to the 
obstacle. Also the boundary condition is $\psi=1$ as $x,y\rightarrow\pm\infty$. We assume 
that for time long enough the system relaxes to a stationary solution, i.e.,  $\psi_t=0$ is satisfied. 
This is well verified for supersonic flow ($M>1.44$) and in 
the following it will be enough to find oblique soliton solutions. Using this condition, 
we express the wave function as $\psi=G/F$ and substitute in eq. (\ref{galileo}). 
Multiplying the resulting equation by $F^3$ one obtains
\begin{equation} 
F[G_{xx}F-2G_xF_x-GF_{xx}+2iM(G_xF-GF_x)+2GF+G_{yy}F-2G_yF_y-GF_{yy}] 
+2GF_xF_x+2GF_yF_y-2|G|^2G=0 , 
\end{equation}
where
the potential $V$ is omitted since we will look for solutions after passing the obstacle\cite{egk-06}.

We now use the well known Hirota techniques. We make the replacements $-GF_{xx}\rightarrow 
+GF_{xx}-2GF_{xx}$ and $-GF_{yy}\rightarrow+GF_{yy}-2GF_{yy}$, multiply by $F$ and rearrange the 
equation as \begin{equation} F^2[(2iMD_x+D^2_x+D^2_y)G.F+2GF]-GF[(D^2_x+D^2_y)F.F+2|G|^2]=0, 
\label{hirota0} \end{equation} where the Hirota $D$-operator is defined generally as $D^n_xf.g 
\equiv (\partial_{x}-\partial_{x'})^ng(x)f(x')|_{x=x'}$.  In our particular case $D_x G.F\equiv 
G_xF-GF_x$, $D^2_x G.F=G_{xx}F-2G_xF_x+GF_{xx}$, $D^2_xF.F=2FF_{xx}-2F_xF_x$.

Equation (\ref{hirota0}) suggests that the system can be put in bilinear form as 
\begin{eqnarray} (2iMD_x+D_x^2+D_y^2)G.F+2GF=\Lambda GF, \label{hirota1} \\ 
(D^2_x+D^2_y)F.F+2GG^*=\Lambda F^2, \label{hirota2} \end{eqnarray} where $\Lambda$ is a constant 
to be determined.
This system of equations have very close similarity to the bilinear form of dark solitons in 
1D-NLS \cite{hirota73,hirotabook}. 

\bigskip

3. {\it Single oblique soliton solution}\newline The single oblique soliton solution was already 
found in \cite{egk-06,annibale2011} assuming null vorticity and using a hydrodynamic formalism. One 
can write it in the stationary frame as 
\begin{equation}\psi=\frac { 
\nu(e^{\xi/2}-e^{-\xi/2})-i\lambda (e^{\xi/2}+e^{-\xi/2}) } {e^{\xi/2}+e^{-\xi/2}} ,
\label{exactss}
\end{equation} 
where $\xi\equiv 2\nu[x\sin\theta-y\cos\theta]$, $\nu\equiv\sqrt{1-\lambda^2}$, 
$\lambda\equiv M\sin\theta$, and $\theta$ is the angle between the soliton and the horizontal axis. 
$M\sin\theta=\pm1$ defines the Mach cone and thus solitons can be found only in the region 
$-\arcsin(1/M)<\theta<\arcsin(1/M)$.

Multiplying the eq.(\ref{exactss}) by an ineffective global phase $i(\lambda+i\nu)$ and numerator and denominator by 
$e^{\xi/2}$ we have \begin{equation} \psi=\frac{1+e^{\xi+2i\alpha}}{1+e^\xi} \,, \end{equation} 
where $e^{i\alpha}\equiv\lambda+i\nu$. One can now readily identify the functions 
\begin{eqnarray} G=1+e^{\xi+2i\alpha} \,,\\ F=1+e^{\xi} . \label{singleFG} \end{eqnarray} After 
substitution of the functions $G$ and $F$ in eqs.(\ref{hirota1},\ref{hirota2}) one finds that 
they remarkably satisfy the bilinear equations provided that $\Lambda=2$. Thus, we were able to show 
that the single soliton solution can be built within Hirota method. It is now natural to look for 
multiple soliton solutions using this formalism. We will pursue this in the following. 

\bigskip

4. {\it Ansatz for two-soliton solution\newline} Based on the similarity of eqs. 
(\ref{hirota1},\ref{hirota2}) with the 1D-NLS 
bilinear form and dark soliton solution \cite{zhang}, we build an ansatz for the two-soliton 
solution in 2D-NLS supersonic flow as \begin{eqnarray} 
G=1+e^{\xi_1+2i\alpha_1}+e^{\xi_2+2i\alpha_2}
   +e^{\xi_1+\xi_2+2i\alpha_1+2i\alpha_2+\varphi_{12}} \,, \label{twoG}\\ 
F=1+e^{\xi_1}+e^{\xi_2}
   +e^{\xi_1+\xi_2+\varphi_{12}} \,, \label{twoF} \end{eqnarray} where 
$\xi_j=2\nu[x\sin\theta_j-y\cos\theta_j]$, $\nu_j=\sqrt{1-\lambda_j^2}$, 
$\exp(i\alpha_j)=\lambda_j+i\nu_j, j=1,2$ and $\varphi_{12}$ is to be determined.

Then, we substitute the ansatz (\ref{twoG},\ref{twoF}) in the equations (\ref{hirota1}) and 
(\ref{hirota2}) and collect terms proportional to $e^0$, $e^{\xi_j}$, $e^{2\xi_j}$, 
$e^{\xi_j+\xi_k}$, $e^{2\xi_j+\xi_k}$, $e^{2\xi_j+2\xi_k}$, $j,k=\{1,2\}$, $j\ne k$. Terms 
proportional to $e^0$, $e^{\xi_j}$, $e^{2\xi_j}$ well satisfy the bilinear equations since they 
correspond to single soliton solutions. The term $e^{2\xi_1+\xi_2}$ gives equations 
\begin{equation} 
e^{2i\alpha_1+\varphi_{12}}\left[ 
4i\nu_2\lambda_2(e^{2i\alpha_2}-1)
+4\nu_2^2(e^{2i\alpha_2}+1)\right]=0 \,, 
\end{equation} 
\begin{equation} 
8\nu_2^2e^{\varphi_{12}}+2e^{\varphi_{12}}(e^{-2i\alpha_2}+e^{2i\alpha_2})-4e^{\varphi_{12}}=0 
\,, \end{equation} which can be shown to be true with little algebra, independently of the value 
of $e^{\varphi_{12}}$. The same for the term $e^{\xi_1+2\xi_2}$. Terms proportional to 
$e^{2\xi_1+2\xi_2}$ give equations that are easily shown to be satisfied.

The remaining terms proportional to $e^{\xi_1+\xi_2}$ generate equations 
\begin{eqnarray} 
4i[(\nu_1\lambda_1-\nu_2\lambda_2)e^{2i\alpha_1}
  +(\nu_2\lambda_2-\nu_1\lambda_1)e^{2i\alpha_2}
  +(\nu_1\lambda_1+\nu_2\lambda_2)(e^{2i\alpha_1+2i\alpha_2}-1)e^{\varphi_{12}}] \nonumber \\ 
+4[S_{-}(e^{2i\alpha_1}+e^{2i\alpha_2})+S_{+}(e^{2i\alpha_1+2i\alpha_2+\varphi_{12}}) 
+S_{+}e^{\varphi_{12}}]=0 \label{primeira} \end{eqnarray} and \begin{equation} 
4(S_{-}+S_{+}e^{\varphi_{12}})+ e^{2i\alpha_1-2i\alpha_2} +e^{-2i\alpha_1+2i\alpha_2} 
+(e^{2i\alpha_1+2i\alpha_2}+e^{-2i\alpha_1-2i\alpha_2})e^{\varphi_{12}}-2-2e^{\varphi_{12}}=0, 
\label{segunda} \end{equation} where $S_{\pm}\equiv\nu_1^2\pm 2\nu_1\nu_2\sigma+\nu_2^2$, 
$\sigma\equiv\sin\theta_1\sin\theta_2+\cos\theta_1\cos\theta_2=\cos(\theta_1-\theta_2)$. Apart 
from the extra factor $\sigma$, these equations are equal to the ones extracted from the 1D-NLS 
\cite{zhang}.

Multiplying the whole equation (\ref{primeira}) by $e^{-i\alpha_1-i\alpha_2}$ and after algebraic 
manipulation we get 
\begin{equation} 
e^{\varphi_{12}}_a=\frac{\sigma^{-1}-\lambda_1\lambda_2-\nu_1\nu_2}
                        {\sigma^{-1}-\lambda_1\lambda_2+\nu_1\nu_2} \label{eqa}
\end{equation} 
and equation (\ref{segunda}) gives 
\begin{equation} 
e^{\varphi_{12}}_b=\frac{\sigma-\lambda_1\lambda_2-\nu_1\nu_2}
                        {\sigma-\lambda_1\lambda_2+\nu_1\nu_2} \label{eqb}, 
\end{equation} 
where the subscripts $a$ and $b$ correspond to $e^{\varphi_{12}}$ extracted from equations 
(\ref{primeira}) and (\ref{segunda}), respectively.

For 1D-NLS, $\sigma$ is equal to 1 and thus $e^{\varphi_{12}}_a=e^{\varphi_{12}}_b$ 
and the two soliton solution is integrable. If $|\theta_1-\theta_2|$ is sufficiently small then 
$\sigma\sim 1$ and the collision of two solitons must be practically elastic. Here the 
treatment is developed regardless the value of $M$. In the symmetrical case, i.e., 
$\theta_1=-\theta_2=\theta$, one can examine the ratio 
$R=e^{\varphi_{12}}_b/e^{\varphi_{12}}_a$, this should provide a measure of how close to Hirota 
integrability is the proposed ansatz. In fig.~1 we show this ratio as a function of $\sin\theta$ 
for different flow velocities $M$. For $M=2$ and small $\theta$ the system is far from 
integrability and the ansatz is poor especially at the collision region and get strong 
deviations from phase shifts. As $M$ is increased the ratio $R$ is closer to 1, independent of the 
angle $\theta$. This was already anticipated in Ref. \cite{annibale2011} using a hydrodynamical 
approach, where it was assumed hipersonic limit $ (M\gg 1) $ so that the equation can be approximated to 
1D-NLS. For the typical case of an impenetrable obstacle with radius $r=1$, solitons are 
generated with $\sin\theta\sim 0.1$ \cite{egk-06}, that gives $R\sim 0.93$ and the collision 
shall appear as almost elastic. This is consistent with the observations in Ref. 
\cite{annibale2011}.

\begin{figure}[bt] \begin{center} \includegraphics[width=8cm]{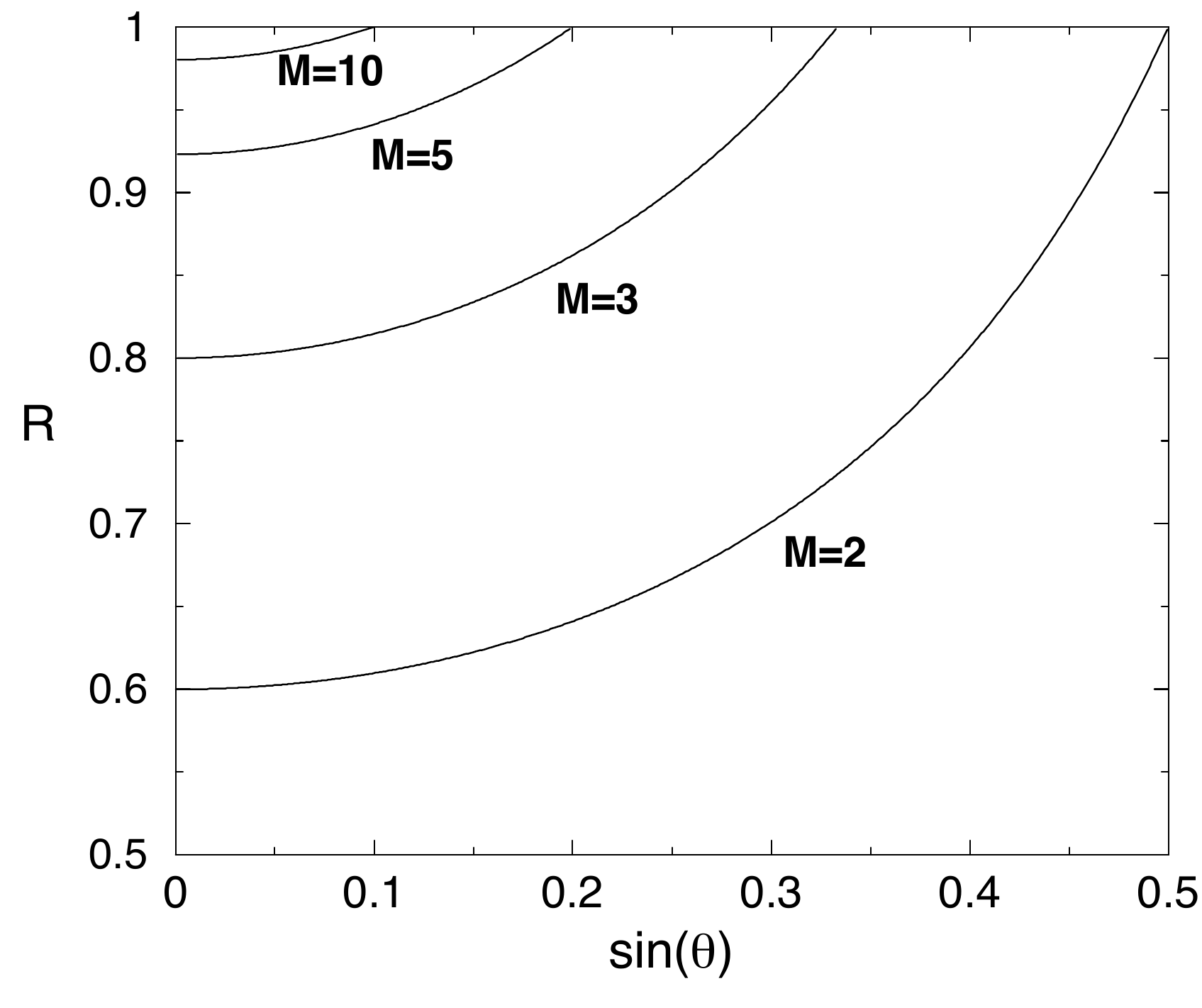} \caption{Results of the 
proposed ansatz similar to the 1D NLS two dark soliton solution. The ratio 
$R=e^{\varphi_{12}}_b/e^{\varphi_{12}}_a$ is between the factors that satisfy each one of the two 
Hirota bilinear equations.} \end{center} \label{fig1} \end{figure}

\bigskip 

5. {\it Phase shifts} \newline
A typical behavior of solitons is that they maintain their shapes after collisions.
However, their positions just after collision are dislocated relative to the free soliton 
propagation. These dislocations are named {\it phase shifts} and are closely related to the factor 
$\exp(\varphi_{12})$. One can calculate the phase shift keeping one of the solitons fixed 
and observing its position when the other soliton is located at infinity\cite{drazin}. 
Following this recipe, we take the ansatz of $\psi=G/F$ given by equations (\ref{twoG},\ref{twoF}), 
multiply numerator and 
denominator by $e^{-\xi_2}$  and take the limit $\xi_2\rightarrow\infty$ giving 
\begin{equation}\psi\sim
e^{2i\alpha_2}\frac 
{ (1+e^{\xi_1+2i\alpha_1+\varphi_{12}}) }
{1+e^{\xi_1+\varphi_{12}} } .
\end{equation}
Thus, comparing with the single soliton solution from eq. (\ref{exactss}), 
the first soliton that depends on $\xi_1$ suffers translation as 
$\xi_1\rightarrow \xi_1+\varphi_{12}$. The dislocation $\delta y$ 
can be calculated simply by 
\begin{equation}
2\nu_1 [x\sin\theta_1-y\cos\theta_1]+\varphi_{12}=
2\nu_1 [x\sin\theta_1-(y+\delta y)\cos\theta_1] 
\end{equation}
giving
\begin{equation}
\delta y=\frac{-\varphi_{12}}{2\nu_1\cos\theta_1} .
\end{equation}
Finally, using expression (\ref{eqa}) in the limit $\sigma\sim 1$ gives
\begin{equation} 
\delta y=\frac{-1}{2\nu_1\cos\theta_1}\ln\left[ \frac 
{1-\lambda_1\lambda_2-\nu_1\nu_2} 
{1-\lambda_1\lambda_2+\nu_1\nu_2} \right] .
\end{equation} 
In the hypersonic limit with $M\gg 1$, $\cos\theta_1$ tends 
to one, and we recover the well known formula of phase shift of 1D dark 
solitons \cite{zs-1973, annibale2011}. Analogous results are obtained for the 
phase shift of the second soliton.

\bigskip

6.{\it Conclusions} \newline We studied the problem of oblique solitons solutions in two-dimensional NLS flow 
using the Hirota bilinear form. This derivation was made in the obstacle frame and assuming stationary flow. 
We were able to build exact single soliton solution with similar form of the 1D-NLS. An ansatz for the 
two-soliton solution was proposed. It is shown that for high Mach number the collision can be considered as 
practically elastic and amplitude and phase can be predicted from NLS-1D approximation, in agreement with 
previous hydrodynamical approach and numerical simulations. Also, solitons with small angles between them 
will collide almost elastically, regardless the velocity $M$. These results are 
relevant for possible experiments like generation of oblique solitons with 
exciton-polaritons that were recently reported in \cite{amo-2010,grosso2011}.

\vspace{1cm}

We thank funding agencies Conselho Nacional de Pesquisa (CNPq) and 
Funda\c{c}\~ao de Amparo \`a Pesquisa do Estado de S\~ao Paulo (FAPESP).


\begin{thebibliography}{99}

\bibitem{egk-06} G. A. El, A. Gammal, and A. M. Kamchatnov, Phys. Rev. Lett. {\bf 97,} 180405 
(2006).

\bibitem{frisch} T. Frisch, Y. Pomeau, and S. Rica, Phys. Rev. Lett. {\bf 69,} 1644 (1992).

\bibitem{berloff2000} N. G. Berloff and P. H. Roberts, J. Phys A: Math. Gen. {\bf 33,} 4025 (2000); 
{\bf 34,} 81 (2001).

\bibitem{rica2001} S. Rica, Physica D {\bf 148,} 221 (2001).

\bibitem{smirnov} V. A. Mironov, A. I. Smirnov, L. A. Smirnov, JETP {\bf 110,} 877 (2010)  
[Zh. \'Eksp. Teor. Fiz. {\bf 137,} 1004 (2010)]. 

\bibitem{winiecki} T. Winiecki, J. F. McCann, and C. S. Adams, Phys. Rev.
Lett. {\bf 82,} 5186 (1999).

\bibitem{ek-06} G. A. El and A. M. Kamchatnov, Phys. Lett. A {\bf 350,} 192 (2006); erratum: Phys.
Lett. A {\bf 352,} 554 (2006).

\bibitem{kgegk-08} E. G. Khamis, A. Gammal, G. A. El, Yu. G. Gladush, and A. M. Kamchatnov, Phys. 
Rev. A {\bf 78,} 013829 (2008).

\bibitem{kp-08} A. M. Kamchatnov and L. P. Pitaevskii, Phys. Rev. Lett. {\bf 100,} 160402 (2008).

\bibitem{kk-11} A. M. Kamchatnov and S. V. Korneev, Phys. Lett. A {\bf 375,} 2577 (2011).

\bibitem{ekkag-09} G. A. El, A. M. Kamchatnov, V. V. Khodorovskii, E. S. Annibale, and A. Gammal, 
Phys. Rev. E {\bf 80,} 046317 (2009).


\bibitem{amo-2010} A. Amo, S. Pigeon, D. Sanvitto, V. G. Sala, R. Hivet, I. Carusotto, F. Pisanello, 
G. Lemenager, R. Houdre, E. Giacobino, C. Ciuti and A. Bramati, Science {\bf 332,} 1167 (2011). 

\bibitem{korneev2011} A. M. Kamchatnov, S.V. Korneev, arXiv:1111.4170 (2011).

\bibitem{grosso2011} G. Grosso, G. Nardin, F. Morier-Genoud, Y. L\'eger,
B. Deveaud-Pl\'edran, Phys. Rev. Lett. {\bf 107,} 245301 (2011).

\bibitem{annibale2011} E. S. Annibale and A. Gammal, Phys. Lett. A {\bf 376,} 46 (2011).

\bibitem{zs-1973} V. E. Zakharov and A. B. Shabat, Sov. Phys. JETP {\bf 37,} 823 (1973) [Zh. \'Eksp. 
Teor. Fiz. {\bf 64,} 1627 (1973)].

\bibitem{hirota73} R. Hirota, J. Math. Phys. {\bf 14,} 805 (1973).

\bibitem {hirotabook} R. Hirota, {\it The direct method in soliton theory}, Cambridge University 
Press, New York, 2004.

\bibitem{zhang} Yi Zhang, Xiao-Na Cai, Cai-Zhen Yao, Hoang-Xian Xu, Mod. Phys. Lett. B {\bf 23}, 
2869 (2009).

\bibitem{drazin} P. G. Drazin and R. S. Johnson, {\it Solitons: An introduction}, Cambridge University 
Press, New York, 1989.
\end{thebibliography}
\end{document}